\documentclass[11pt]{article}
\usepackage{geometry}
\usepackage[utf8x]{inputenc}
\usepackage[english]{babel}
\usepackage{microtype}
\usepackage{framed}
\usepackage{xcolor}
\usepackage{authblk}
\usepackage{comment}
\usepackage{graphicx}
\usepackage{booktabs}
\usepackage{csquotes}
\usepackage{multicol}
\usepackage{flushend}
\usepackage[hybrid]{markdown}
\usepackage[graphicx]{realboxes}
\usepackage[binary-units]{siunitx}

\usepackage{listings}
\PrerenderUnicode{í}
\PrerenderUnicode{ě}

\newcommand\instance[1]{\texttt{#1}}

\usepackage[unicode]{hyperref}
\hypersetup{pdftitle={Duet Benchmarking: Improving Measurement Accuracy in the Cloud (Accepted Preprint Version)}}
\hypersetup{pdfauthor={Lubomir Bulej, Vojtech Horky, Petr Tuma, Francois Farquet and Aleksandar Prokopec}}

\title{{Duet Benchmarking: \\ Improving Measurement Accuracy in the Cloud \\ (Accepted Preprint Version)}}
\author[1]{Lubomír Bulej}
\author[1]{Vojtěch Horký}
\author[1]{Petr Tůma}
\author[2]{François Farquet}
\author[2]{Aleksandar Prokopec}
\affil[1]{
    Charles University

    Faculty of Mathematics and Physics

    Department of Distributed and Dependable Systems

    Prague, Czech Republic

    \texttt{\{name.surname\}@d3s.mff.cuni.cz}

    ~
}
\affil[2]{
    Oracle Labs

    Zurich, Switzerland

    \texttt{\{name.surname\}@oracle.com}
}

\begin{document}

\maketitle

\begin{abstract}
We investigate the duet measurement procedure,
which helps improve the accuracy of performance comparison experiments conducted on shared machines
by executing the measured artifacts in parallel and evaluating their relative performance together, rather than individually.
Specifically, we analyze the behavior of the procedure in multiple cloud environments and use experimental evidence
to answer multiple research questions concerning the assumption underlying the procedure. We demonstrate improvements
in accuracy ranging from $2.3\times$ to $12.5\times$ ($5.03\times$ on average) for the tested ScalaBench (and DaCapo)
workloads, and from $23.8\times$ to $82.4\times$ ($37.4\times$ on average) for the SPEC CPU 2017 workloads.
\end{abstract}

\maketitle

\section* {Rights}

\noindent Uploaded to ArXiV under the ACM Copyright Policy Version 9.
Copyright 2020 ACM and the authors.
This is the author version of the work.
Posted for your personal use.
Not for redistribution.
The definitive Version of Record was published in \emph{Proceedings of the 2020 ACM/SPEC International Conference on
Performance Engineering (ICPE ’20), April 20–24, 2020, Edmonton, AB, Canada}, \url{https://doi.org/10.1145/3358960.3379132}.

\medskip

\noindent This work was partially supported by the ECSEL Joint Undertaking (JU) under grant agreement No 783162
and the Charles University institutional funding (SVV).

\section{Introduction}

At the heart of various performance comparison activities is a measurement experiment,
whose statistical nature involves an inherent trade off between execution time
and sensitivity to differences in performance.
Longer experiment times average over noise in the measurement data and provide more accurate results,
but are also expensive both in terms of time and computing resources. Conversely, shorter
execution times may cause the loss of sensitivity or report false alarms.
This is a problem when automating performance test execution and evaluation
\cite{HuangPerformanceRegressionTesting2014, OliveiraPerphecyPerformanceRegression2017}.

Importantly, the resource requirements for performance testing are not constant, but rather
reflect the development activities, the test scenarios, and the desired level of sensitivity.
To satisfy the changing resource requirements, it is therefore attractive to consider offloading the performance testing activities to the cloud.

\begin{figure}
\includegraphics[width=\linewidth]{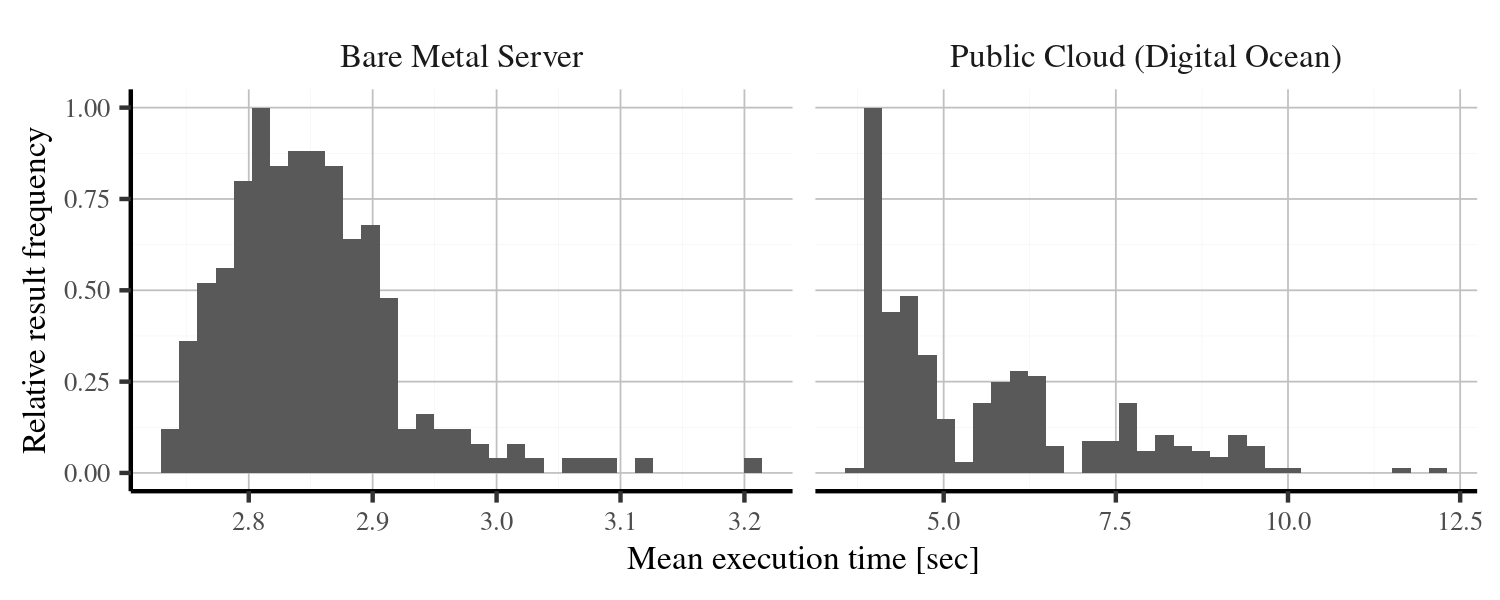}
\caption{Distribution of observed mean execution times of the \textsf{avrora} benchmark, running on an otherwise idle bare-metal server and on a public cloud machine.
Note the min-max range, which is about \SI{16}{\percent} of the mean on the bare-metal server and about \SI{150}{\percent} in the cloud.}
\label{fig:cloud-is-noisy}
\end{figure}

A specific hurdle in this context is the fact that the cloud does not necessarily provide the performance stability required for performance testing.
Performance measurements in the cloud are noisy, in part due to lack of control over hardware configuration, in part due to overhead of virtualization,
but most importantly due to interference from colocated workloads of other tenants \cite{IosupPerformanceVariabilityProduction2011, LeitnerPatternsChaosStudy2016, LaaberSoftwareMicrobenchmarkingCloud2019}.
To illustrate this, Figure~\ref{fig:cloud-is-noisy} shows the distribution of mean task execution times
for iterations of an example benchmark from the DaCapo suite, both on a bare-metal server and
on a virtual machine running in a public cloud.

Our earlier work~\cite{BulejInitialExperiments2019} introduced the idea of the \emph{duet measurement procedure},
which improves measurement accuracy in shared resource environments, such as virtual machine instances in the cloud.
The procedure is based on the assumption that performance fluctuations due to interference tend to impact similar tenants equally,
and attempts to maximize the likelihood of such equal impact by executing the measured artifacts in parallel.
The subsequent computation filters out the fluctuations by considering the relative performance
of the measured artifacts together.

The assumptions of the duet measurement procedure hinge on detailed technical properties of both the measurement platform and the executing workloads.
In the cloud, such properties typically cannot be controlled or guaranteed, we therefore subject the procedure to a thorough experimental evaluation
with the goal of analyzing the overall behavior and documenting the observed accuracy.
Based on experimental evidence, we answer specific research questions concerning the assumption
underlying the procedure and explain the technical mechanisms behind the observations:

\begin{itemize}
\item We demonstrate improvements in accuracy that range from $2.3\times$ to $12.5\times$ ($5.03\times$ on average) for the tested ScalaBench~\cite{SeweCapoConScala2011} (and DaCapo~\cite{BlackburnDaCapoBenchmarksJava2006}) workloads, and from $23.8\times$ to $82.4\times$ ($37.4\times$ on average) for SPEC CPU 2017 workloads~\cite{softwareSpecCpu2017}.
\item We show that the accuracy improvements are due to the ability of the duet procedure to isolate synchronized interference, and that this interference arises with resource sharing.
\item We evaluate how the specific patterns of concurrent execution and uneven resource utilization impact the ability of the duet procedure to measure performance differences.
\end{itemize}

As an essential overall contribution, our results indicate that cloud-based virtual machines can provide a viable platform for conducting an entire class of performance testing experiments based on comparing task execution times of benchmark workloads.

Section~\ref{sec:background} provides additional background and motivation for performance regression testing as our specific application context.
Section~\ref{sec:approach} presents an overview of the duet measurement procedure and the associated computations.
Section~\ref{sec:evaluation} presents experimental evaluation answering specific research questions
that naturally arise when using duet measurements and observing the effects on measurement accuracy.
We review related work in Section~\ref{sec:related} and conclude the paper in Section~\ref{sec:conclusion}.

\section{Background and Motivation}
\label{sec:background}

The motivation for our work is performance regression testing, that is,
the task of detecting performance changes between two versions of a software project.
To this end, we use benchmark workloads to exercise both versions of the project,
measuring and comparing task execution times of individual workloads between the two versions.

Essential to performance regression testing is robust performance change detection.
The task execution times observed on a real system are influenced by different sources of variability at different levels of granularity -- the comparison therefore relies on statistical hypothesis testing to accommodate the inherent variability in the data, and the performance testing procedure must ensure that significant sources of variability are sufficiently represented in the data \cite{GeorgesStatisticallyRigorousJava2007, BulejUnitTestingPerformance2016, AbediConductingRepeatableExperiments2017}.

To provide sufficient variability, benchmarks repeatedly execute the same task (in a single process) and measure the task execution time in each \emph{iteration}.
This captures variability caused by factors that can manifest at any time during benchmark execution,
and which can influence the execution time of any iteration, such as scheduling, memory caches, or background load.
In addition, benchmarks are executed repeatedly to obtain execution times from multiple benchmark \emph{runs} (in multiple processes).
This captures variability caused by factors that can change between runs, but rarely change within a single run,
such as process memory layout, or decisions of managed platforms such as the Java Virtual Machine.

As a general rule, the variability in the observed execution times determines the magnitude of performance changes
that can be reliably detected in a given time, or alternatively, the time needed to detect performance changes
of a given magnitude. For a quick illustration of the computational resources needed for performance
regression testing, we use the open source GraalVM project~\cite{graalvm-repo}, where the developers
contribute on average 5 merge commits per day and want to test these commits for performance changes
on a selection of 60 workloads from multiple benchmark suites.
When using Java workloads for tests at the \SI{99}{\percent} confidence level, we can realistically assume to need data from 30 benchmark runs,
each executing for 10 minutes (to get past some of the warm up effects). This sums up to 10 machine hours for a single experiment involving
one version pair and one benchmark, and becomes 3000 machine hours per day for all experiments, which is an overwhelming figure.

To pare down the resource demands, we can limit the amount of testing actually
done~\cite{HegerAutomatedRootCause2013, OliveiraPerphecyPerformanceRegression2017},
however, that alone may not solve the problem of infrastructure capacity limits.
This is where cloud resources come into consideration, yet it is unclear if they are of any use for performance regression testing -- the degree of control over the experimental platform, which allows obtaining accurate measurements on the local infrastructure, is not available in the cloud.
Furthermore, cloud providers offer abstract virtual machine types that can run on different
types of physical hosts~\cite{LeitnerPatternsChaosStudy2016}, resulting in different execution times even for the same code.
Finally, cloud virtual machines suffer from performance interference of neighbor workloads, which the virtualization technology cannot entirely eliminate.
This also holds for continuous integration solutions executing in the cloud, such as Travis~\cite{softwareTravis} or GitLab Runner~\cite{softwareGitlabRunner}.

In summary, we need a procedure that takes the characteristics of the cloud into account and makes it useful for performance testing,
even if it only allows to quickly process many versions and flag suspect cases for more thorough measurements on dedicated infrastructure.

\section{Duet Measurement Procedure}
\label{sec:approach}

Measurements in the cloud are subject to performance interference, which manifests as noise that may randomly affect any measured data.
To account for the probabilistic nature of the interference, we have to repeat the measured operation enough times to obtain
a representative sample of measurements, and then calculate confidence intervals for any values derived from the measurements.
In experiments involving multiple workloads there is a risk of a systematic bias in the measured data if the probability
of a workload being influenced by interference is not equal for all workloads.
The current best practice uses randomized interleaving of workloads~\cite{AbediConductingRepeatableExperiments2017},
which---for a long enough experiment---avoids the bias by equalizing the probability of interference for all workloads.

The \emph{duet measurement procedure} also avoids bias by equalizing probability of interference,
but is specifically tailored for experiments comparing performance of two (related) workloads.
The two workloads are executed in parallel, inside a virtual machine with two virtual cores,
with each workload restricted to one virtual core. The workloads are synchronized using
a shared memory barrier, so that their measured operations always start at the same time.
This setting ensures that any external interference on the virtual machine impacts both workloads simultaneously,
which equalizes the probability of interference between the workloads for each paired measurement and thus
avoids the bias immediately---rather than only for a long enough experiment.

We derive the confidence interval for the ratio of task execution times, which describes the relative performance of the two workloads,
using a Monte Carlo procedure based on standard bootstrap confidence interval computation~\cite{HesterbergWhatTeachersShould2014},
explained in detail in ~\cite{BulejInitialExperiments2019}:

\begin{enumerate}
    \item For a pair of workloads $x$ and $y$ and an experiment with $R$ runs of $I$ iterations each, we denote $x_{r,i}$ and $y_{r,i}$ the task execution times of the respective workloads, measured in iteration $i \in 1 \ldots I$ of run $r \in 1 \ldots R$.

    \item For each $r$ and $i$, we use the paired samples $x_{r,i}$ and $y_{r,i}$ to calculate the corresponding (speedup) sample $s_{r,i}$ of the ratio between task execution times of workloads $x$ and $y$:
    \begin{center}
    $\forall r \in 1 \ldots R, \forall i \in 1 \ldots I: s_{r,i} = {x_{r,i} \over y_{r,i}}$
    \end{center}

    \item For each run, we aggregate the speedup samples across iterations in a run by computing the geometric mean:
    \begin{center}
    $\forall r \in 1 \ldots R: gms_r = \sqrt[I]{s_{r,1} \cdot s_{r,2} \ldots s_{r,I}}$
    \end{center}

    \item We aggregate the geometric means across all runs in an experiment by computing the grand geometric mean:
    \begin{center}
    $ggms = \sqrt[R]{gms_1 \cdot gms_2 \ldots gms_R}$
    \end{center}
    The value $ggms$ represents a point estimate of the ratio of task execution times between workloads $x$ and $y$, i.e., the relative performance of the two workloads.\label{step:ggms}

    \item We use non-parametric bootstrap to estimate the percentile confidence interval for $ggms$, drawing with replacement from $gms_\bullet$ and computing $ggms^*$ (step~\ref{step:ggms} applied on the sample drawn from $gms_\bullet$) as Monte Carlo estimates for $ggms$.
\end{enumerate}

\noindent
When the confidence interval for $ggms$ (mean ratio of task execution times) straddles $1.0$, we consider the observed performance of the two workloads equal, otherwise we report a performance difference.

\section{Experimental Evaluation}
\label{sec:evaluation}

We examine the duet measurements using multiple experiments designed to answer specific research questions.
Before introducing the research questions and the experiments, we outline the experimental environment.
For detailed information, please consult the appendix.

The duet measurements target shared resource environments common in clouds, most of our measurements therefore execute in clouds.
As the main cloud platform, we use the Amazon Elastic Cloud, specifically the \instance{t3.medium}, \instance{t3a.medium},
\instance{m5.large} and \instance{m5a.large} instance types.
As our second cloud platform, we use the Travis CI infrastructure~\cite{softwareTravis},
which in turn uses otherwise unspecified Google Compute Engine platform machine instances.
As our third cloud platform, we use the GitLab CI infrastructure~\cite{softwareGitlabRunner}
backed by Digital Ocean machine instances.
In addition to the three public cloud platforms, we carry out measurements on a private cloud running the Proxmox Virtual Environment.
Finally, we run bare metal measurements that are to represent the most stable baseline for comparison.

To approximate realistic workloads, we use benchmark suites -- SPEC CPU 2017~\cite{softwareSpecCpu2017} for statically compiled and optimized workloads, and ScalaBench~\cite{SeweCapoConScala2011} (with DaCapo~\cite{BlackburnDaCapoBenchmarksJava2006}) for dynamically compiled and optimized workloads.
From SPEC CPU 2017, we execute the rate workload variants (23 workloads in total). From ScalaBench and DaCapo, we execute all workloads
except \textsf{actors}, \textsf{batik}, \textsf{eclipse}, \textsf{tomcat}, \textsf{tradebeans} and \textsf{tradesoap}, which fail for various reasons (20 workloads in total).
We use the OpenJDK 1.8.0 JVM, run with fixed heap size and disabled garbage collector ergonomics,
other virtual machine settings were left at their defaults.

To provide information on result variance, we execute all benchmarks multiple times
(on average over 20 runs for each workload on the Amazon \instance{t} instances,
over 40 runs on the Amazon \instance{m} instances,
and over 100 runs on the other platforms),
and use random samples of 10 runs for all computations.
On the faster execution platforms (public cloud at full speed, private cloud, bare metal),
we collect the timing of the first 100 iterations or 10 first minutes of execution within each run, whichever comes first.
On the slower execution platforms (public clouds with token bucket processor allocation), it is 100 iterations or 60 minutes.
We do not execute the SPEC CPU 2017 workloads on the Amazon \instance{t} instances and on the Travis CI infrastructure,
because both lack the computing power to execute the benchmark in reasonable time.
For the SPEC CPU 2017 workloads, which exhibit virtually no startup artifacts, we use the timing of all iterations.
For the ScalaBench workloads, which exhibit startup artifacts related to dynamic compilation, we discard the timing of the first half of iterations.
We apply outlier filtering with winsorization in all computations, replacing at most one observation in a run with its nearest neighbor
when that observation is further than 20\% away from the min-max range of the remaining observations.
Our bootstrap computations use 10000 replicates.

The constants above were determined by informal experiments to provide reasonable measurement time and reasonable stability across the workload spectrum.
In an actual performance testing environment, the numbers would be chosen per platform and per workload
using established procedures such as~\cite{HeStatisticsBasedPerformanceTesting2019, MaricqTamingPerformanceVariability2018},
however, introducing this practice here would prevent us from comparing different measurement procedures under similar conditions.

\subsection{RQ1: Accuracy Improvements}

The very purpose of the duet procedure is to improve the accuracy of performance comparison experiments.
Our first research question directly addresses this purpose:
\emph{Are the performance comparisons made with the duet procedure more accurate than performance comparisons done using standard methods ? (RQ1)}

The standard way to express the measurement accuracy is to treat the individual measurements as observations of a random variable with an unknown parameter of interest, such as the mean value. The goal of the measurement is to estimate this unknown parameter, and the accuracy of this estimate characterizes the overall measurement accuracy. An intuitive way to present the accuracy of the estimate, which we also use in this paper, is with confidence intervals~\cite{HesterbergWhatTeachersShould2014}. For the duet measurements, we use the 99\% confidence intervals for the mean of ratios computed with the procedure in Section~\ref{sec:approach}. As a representative standard method that we compare against, we use the common 99\% bootstrap confidence intervals for the difference of means, computed using the procedure in~\cite{BulejUnitTestingPerformance2016}, with random measurement interleaving, as recommended in~\cite{AbediConductingRepeatableExperiments2017}.

We collect the accuracy information using A/A measurements, that is,
we compare two sets of measurements that use the same workload and the same instance type.
For each workload and instance type, the comparison gives us two confidence intervals,
one for the mean ratio of the workload execution times computed using the duet procedure, and
one for the difference of the mean workload execution times computed using the standard method.
By construction of the experiment, the two intervals must respectively straddle $1.0$ and $0.0$,
and the width of the two intervals expresses the accuracy achieved by the two procedures.

A direct comparison of the two confidence intervals is hindered by the fact that
the intervals produced by the duet procedure are centered around $1.0$ but
the intervals produced by the standard method are centered around $0.0$.
We therefore convert both types of confidence intervals to a value expressing their width relative to mean performance -- for the mean of ratios interval $(ggms_{lo}, ggms_{hi})$ we report $ggms_{hi} - ggms_{lo}$, and for a difference of means interval $(\textit{di{f}f}_{lo}, \textit{di{f}f}_{hi})$ we report $(\textit{di{f}f}_{hi} - \textit{di{f}f}_{lo}) / mean$, where $mean$ is the sample mean computed from all samples (all samples concern the same workload and can therefore be averaged).

\begin{figure*}[t!]
\includegraphics[width=\linewidth]{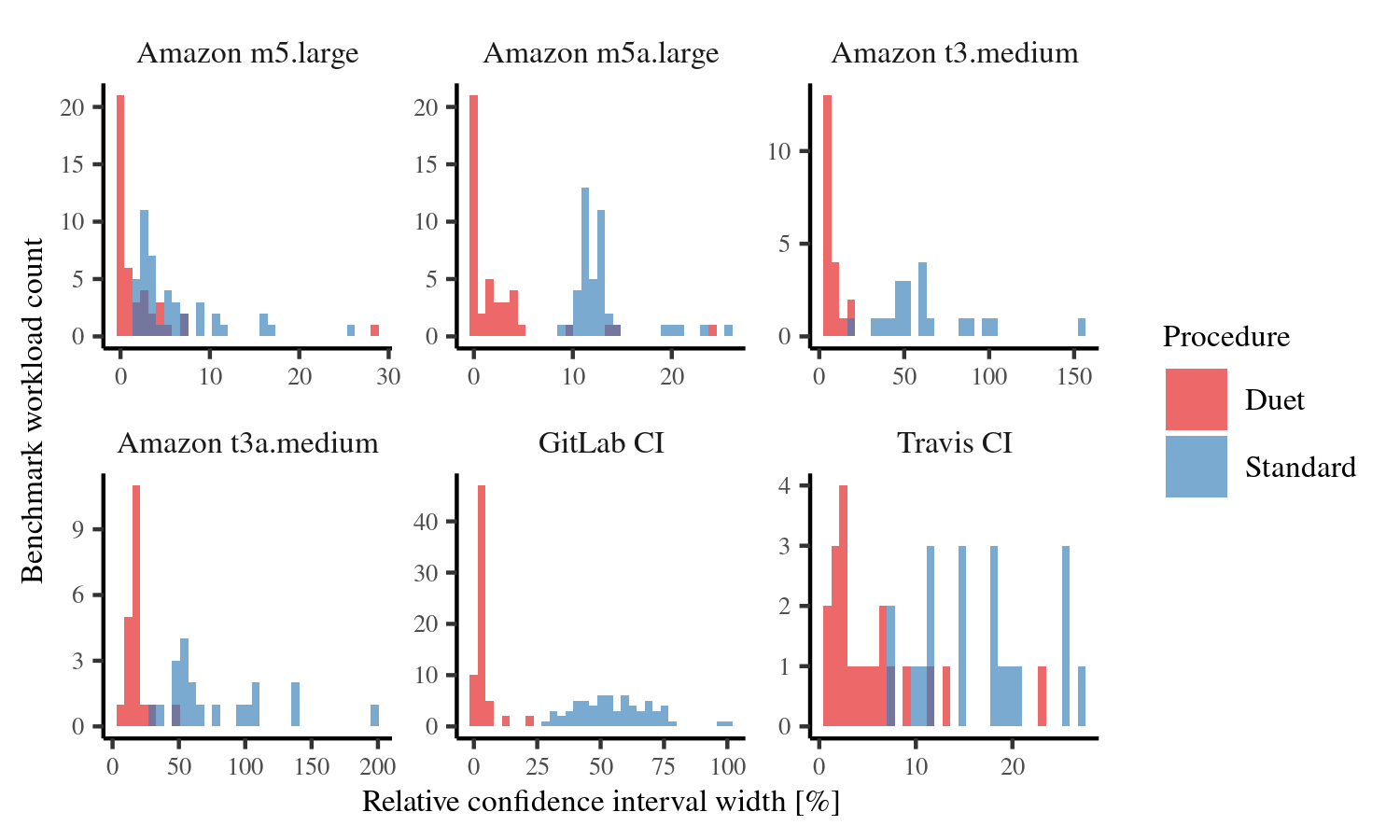}
\caption{Accuracy expressed as relative 99\% confidence interval width, 10 runs, aggregated across all workloads.}
\label{fig:accuracy-public-cloud}
\end{figure*}

\begin{table}[hbt!]
\caption{Average reduction in relative 99\% confidence interval width from the standard procedure to the duet procedure, geomean.}
\label{tbl:accuracy-public-cloud}
\begin{center}
\begin{tabular}{lcc}
\toprule
Platform & ScalaBench & SPEC CPU 2017 \\
\midrule
Amazon m5.large & $ 2.3\times$ & $26.6\times$ \\
Amazon m5a.large & $3.86\times$ & $82.4\times$ \\
Amazon t3.medium & $9.13\times$ & --- \\
Amazon t3a.medium & $3.99\times$ & --- \\
GitLab CI & $12.5\times$ & $23.8\times$ \\
Travis CI & $3.97\times$ & --- \\
\midrule
Average & $5.03\times$ & $37.4\times$ \\
\bottomrule
\end{tabular}
\end{center}
\end{table}

\begin{figure}[hbt!]
\includegraphics[width=\linewidth]{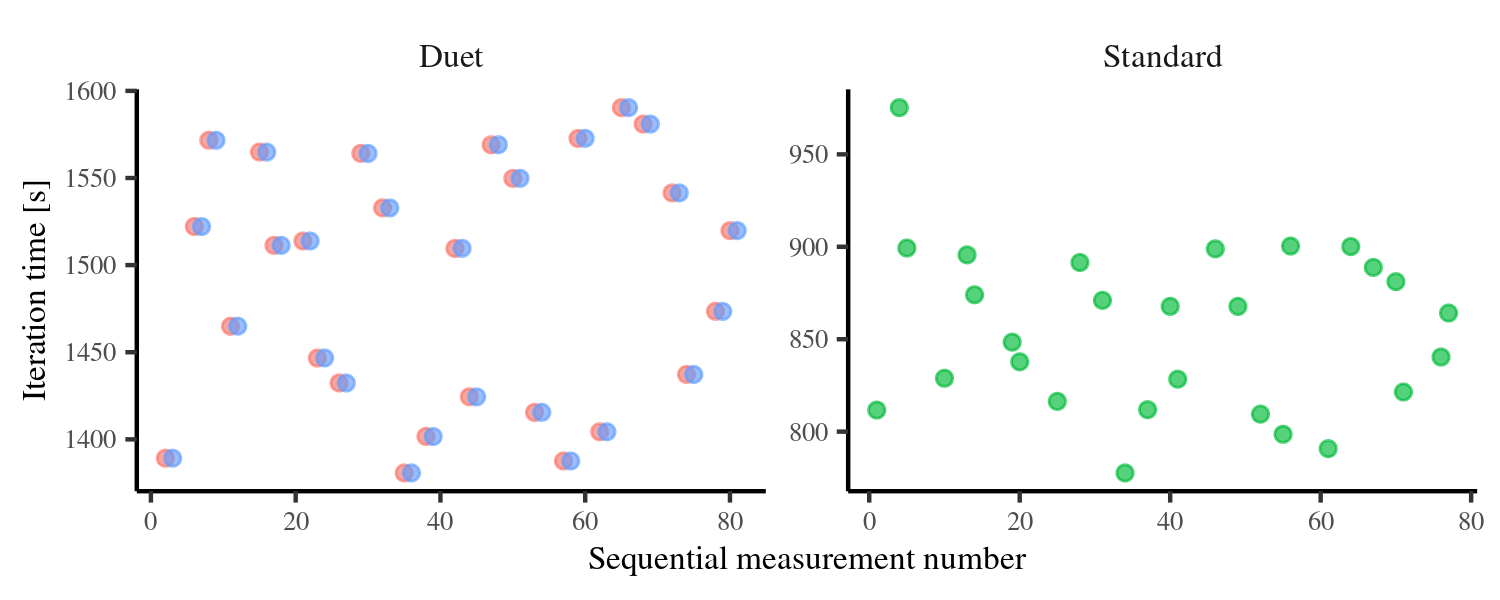}
\caption{Individual measurement samples for the 503.bwaves\_r workload on the Amazon \instance{m5a.large} platform.
Colors in the duet procedure distinguish samples collected in parallel.}
\label{fig:pairwise-accuracy-example}
\end{figure}

Figure~\ref{fig:accuracy-public-cloud} shows the distribution of the \SI{99}{\percent} confidence interval widths
on the public cloud platforms, aggregated across all workloads.\footnote{We use the \SI{99}{\percent} confidence
level throughout the presentation, however, other confidence levels provide reasonably similar results.}
The distribution indicates that the duet procedure generally delivers more narrow confidence intervals and therefore better accuracy.
Table~\ref{tbl:accuracy-public-cloud} aggregates the improvement in accuracy for each platform and benchmark,
expressed as the average reduction of the relative confidence interval width.
For the ScalaBench workloads, the duet procedure computes on average $5.03$ times more narrow intervals than the standard method.
For the SPEC CPU 2017 workloads, the duet procedure computes on average $37.4$ times more narrow intervals,
in part because the workloads are much more stable and even small measurement fluctuations
due to resource sharing are therefore more significant.
Figure~\ref{fig:pairwise-accuracy-example} provides more insight into this behavior by plotting the individual measurement samples for both the duet procedure and the standard method on one arbitrarily selected workload and platform combination.
While the measurement fluctuations are always present, the samples collected in parallel by the duet procedure move (vertically) very much in tandem, almost perfectly matching the assumptions of the duet procedure.

To give an intuitive illustration of the improvement in accuracy, we look at the associated measurement costs.
The mean confidence intervals tend to shrink with the square root of the sample counts -- asymptotically, this holds due to the Central Limit Theorem, but here we refer rather to empirical observations at small sample counts, where we see similar behavior.
A twofold improvement in accuracy at constant sample count therefore roughly corresponds to a fourfold reduction in sample count at constant accuracy.
Note that the measurement costs are also impacted by different platform requirements -- where the standard method requires sufficient resources to run a single workload copy, the duet procedure requires resources for two workloads executing concurrently.

\subsection{RQ2: Synchronized Interference}

At the core of the duet procedure is the idea to expose the compared workloads to the same interference.
To achieve that, the procedure modifies the way the workloads are executed and the way the results are processed.
We therefore need to determine whether the observed accuracy improvements are due to the synchronized interference,
rather than a side effect of the modifications in workload execution and results processing.
\emph{Can we attribute the improved accuracy exhibited by the duet procedure to both workloads suffering from synchronized interference ? (RQ2)}

To isolate the contribution of synchronized interference from the other modifications introduced by the duet procedure,
we use the existing measurements, but adjust the confidence interval computation from Section~\ref{sec:approach}.
Where the duet procedure normally computes ratios from measurements collected at the same time,
we now perform a random shuffle and use ratios from unrelated measurements.
That way, we preserve all other aspects of the duet procedure, but obtain results that do not benefit from synchronized interference.

\begin{figure*}[t!]
\includegraphics[width=\linewidth]{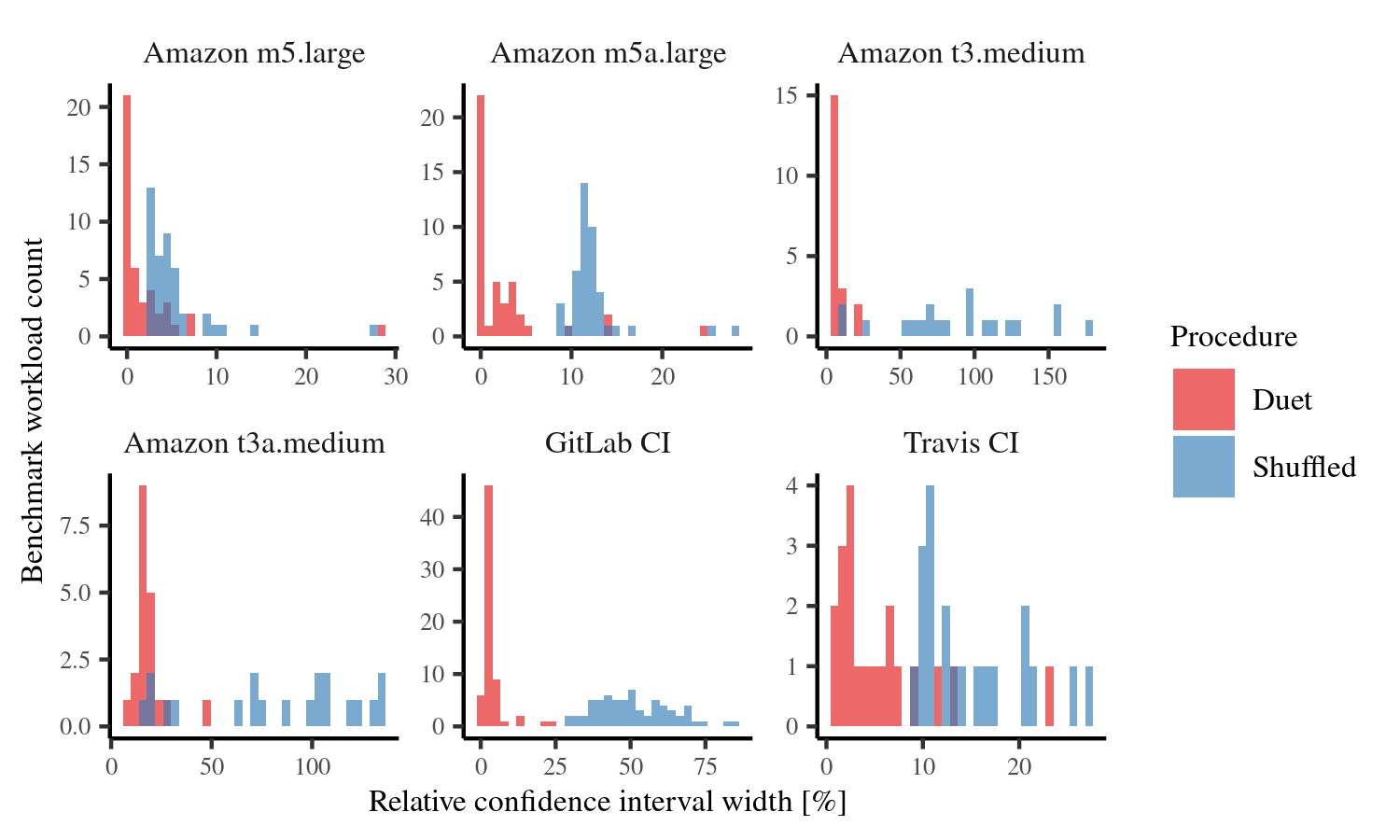}
\caption{Impact of random shuffling on relative 99\% confidence interval width, 10 runs, aggregated across all workloads.}
\label{fig:shuffled-public-cloud}
\end{figure*}

Figure~\ref{fig:shuffled-public-cloud} shows the impact of shuffling on the distribution of the confidence interval widths.
The distribution demonstrates that the duet procedure indeed benefits particularly from synchronized interference.
We can also note that the confidence interval widths obtained with shuffling are very similar
to the confidence interval widths from Figure~\ref{fig:accuracy-public-cloud}
computed by the standard method.
If we compute the aggregate improvement in accuracy after shuffling -- an analogue of Table~\ref{tbl:accuracy-public-cloud} but without synchronized interference -- we obtain a total of $1.02$ for the ScalaBench workloads and $1.03$ for the SPEC CPU 2017 workloads, suggesting not only that the ability to deal with synchronized interference is the major factor contributing to improved accuracy, but also that other factors inherent to the duet procedure, such as the concurrent workload execution, are not a major detriment.

\subsection{RQ3: Resource Sharing}

The third aspect of the duet procedure we investigate is whether
the presence of synchronized interference is due to resource sharing common in clouds,
or whether some other property of our experiments may account for the observed behavior.
\emph{Is the presence of synchronized interference associated with the existence of other workloads that share the same computing platform ? (RQ3)}

The only way to control other workloads on the same platform in the public could is to rent an entire physical machine,
however, that option also removes the virtualization infrastructure, making apples-to-apples comparison impossible.
Instead, we therefore use private cloud measurements and control the utilization of the physical servers
backing the virtual machine instances.
In one set of measurements, we make sure each physical server runs only the measured workload.
In the other set of measurements, we add a competing workload with the potential to saturate the physical server.
Our competing workload is the composite configuration of the SPEC JBB 2015 benchmark,
which generates a variable workload pattern across all cores of the physical server,
moving between zero and peak utilization with a period of about 150 minutes.
The workload approximates an enterprise business application and is therefore relevant in the cloud context.

\begin{figure}[hbt!]
\includegraphics[width=\linewidth]{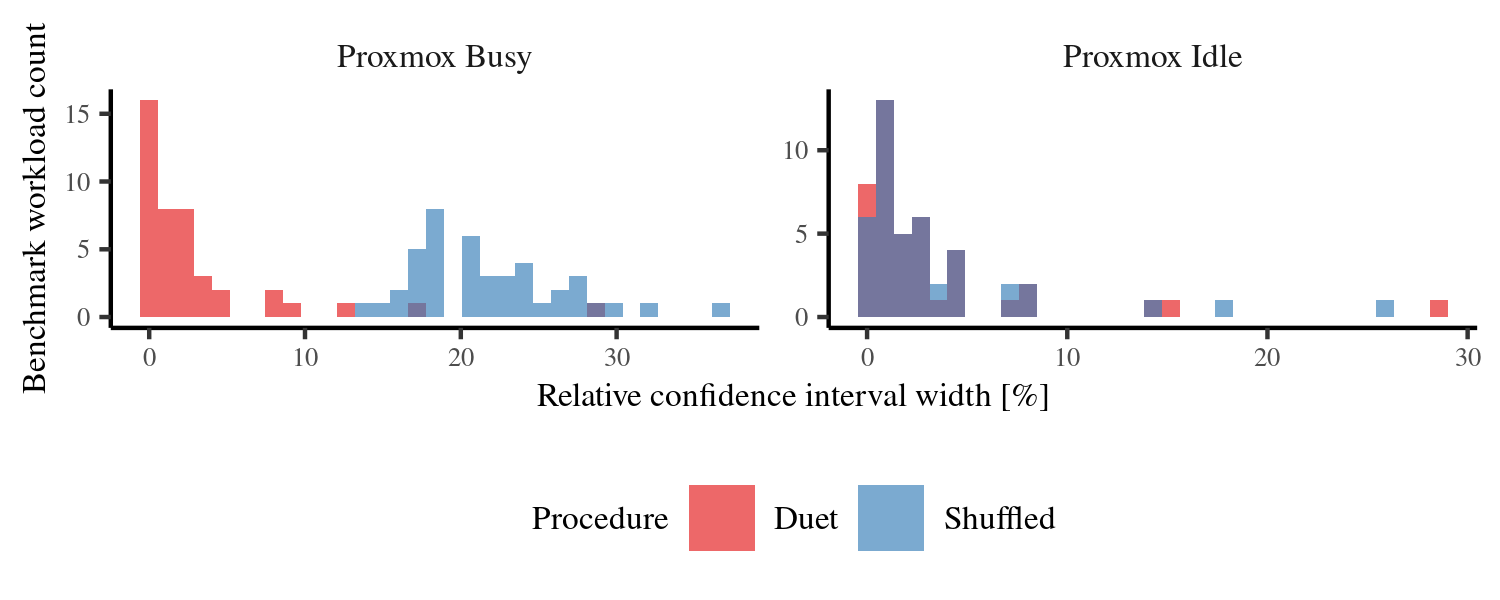}
\caption{Impact of resource sharing on random shuffling in private cloud, idle vs busy with competing workload, expressed as relative 99\% confidence interval width, 10 runs, aggregated across all workloads.}
\label{fig:sharing-private-cloud}
\end{figure}

Figure~\ref{fig:sharing-private-cloud} demonstrates the impact of resource sharing on confidence intervals,
again computed using either ratios from measurements collected at the same time,
or ratios from unrelated measurements after a random shuffle.
In the left-hand part of the plot, where the measurements were performed with resource contention, shuffling changes the confidence intervals significantly.
In the right-hand part of the plot, where the measurements were without resource contention, shuffling has almost no effect.
This confirms our hypothesis that the synchronized interference we observe and tackle with the duet procedure
is indeed due to resource sharing.

\subsection{RQ4: Measuring Differences}

The duet procedure does not always utilize the computing resources evenly.
Assume A/B measurements where the duet workloads differ in length, with A shorter and B longer.
The concurrent workload execution phase, as long as A, will be followed by
an isolated workload execution phase, as long as the remaining part of B.
This makes the execution conditions for the two workloads differ --
while A always competes for the shared resources,
B executes partially with and partially without
such competition.
It may therefore finish faster than if the computing resources were utilized evenly,
making the duet procedure underestimate the workload execution time ratio.

An underestimated workload execution time ratio is not necessarily a serious issue.
Our motivation is the ability to detect performance changes during regression testing.
In this context, it is enough to use the cloud to reliably detect the presence of a change,
additional measurements to assess the magnitude can be performed in a controlled environment.
We should, however, still seek to understand the impact of uneven resource on the measurements.
\emph{How does uneven resource utilization impact the estimated workload execution time ratio ? (RQ4)}

We answer the research question by arranging workloads with known execution time ratio in an A/B measurement and looking at the actual ratio measured and reported by the duet procedure. We do this first in the private cloud, where we have more control over the workload duration and resource utilization, and next in the public cloud, where we can use previous measurements.

\medskip\noindent\textbf{Private cloud.}
To get sufficient control over workload duration and resource utilization, we move from the benchmarks to four entirely artificial workloads, designed to utilize a given resource for a given operation count.
We refer to the four workloads as \emph{integer} (an integer loop running entirely from level 1 caches), \emph{float} (a floating point computation also running entirely from level 1 caches), \emph{cache} (a linear memory walk over \SI{4}{\mebi\byte} of data that mostly hits in the last level cache), and \emph{memory} (a random memory walk over \SI{64}{\mebi\byte} of data that mostly misses in the last level cache).
The \emph{integer} and \emph{float} workloads are sensitive mostly to hyperthreading and power management,
while the \emph{cache} and \emph{memory} workloads add sensitivity to competition on the memory resources.

We first calibrate the artificial workloads on the private cloud platform,
obtaining operation counts that yield roughly \SI{100}{\milli\second} executions.
For each artificial workload, we then execute A/B measurements where
A executes the workload using the calibrated operation count and
B executes the same workload using twice the count of A.
For the artificial workload, the operation count translates directly into execution time,
we would therefore desire to observe iteration times with the ratio of $2.0$.\footnote{
Note that the relationship between operation count and execution time does not hold
for the benchmark workloads, one reason why artificial workloads are used here.
}

\begin{figure}[hbt!]
\includegraphics[width=\linewidth]{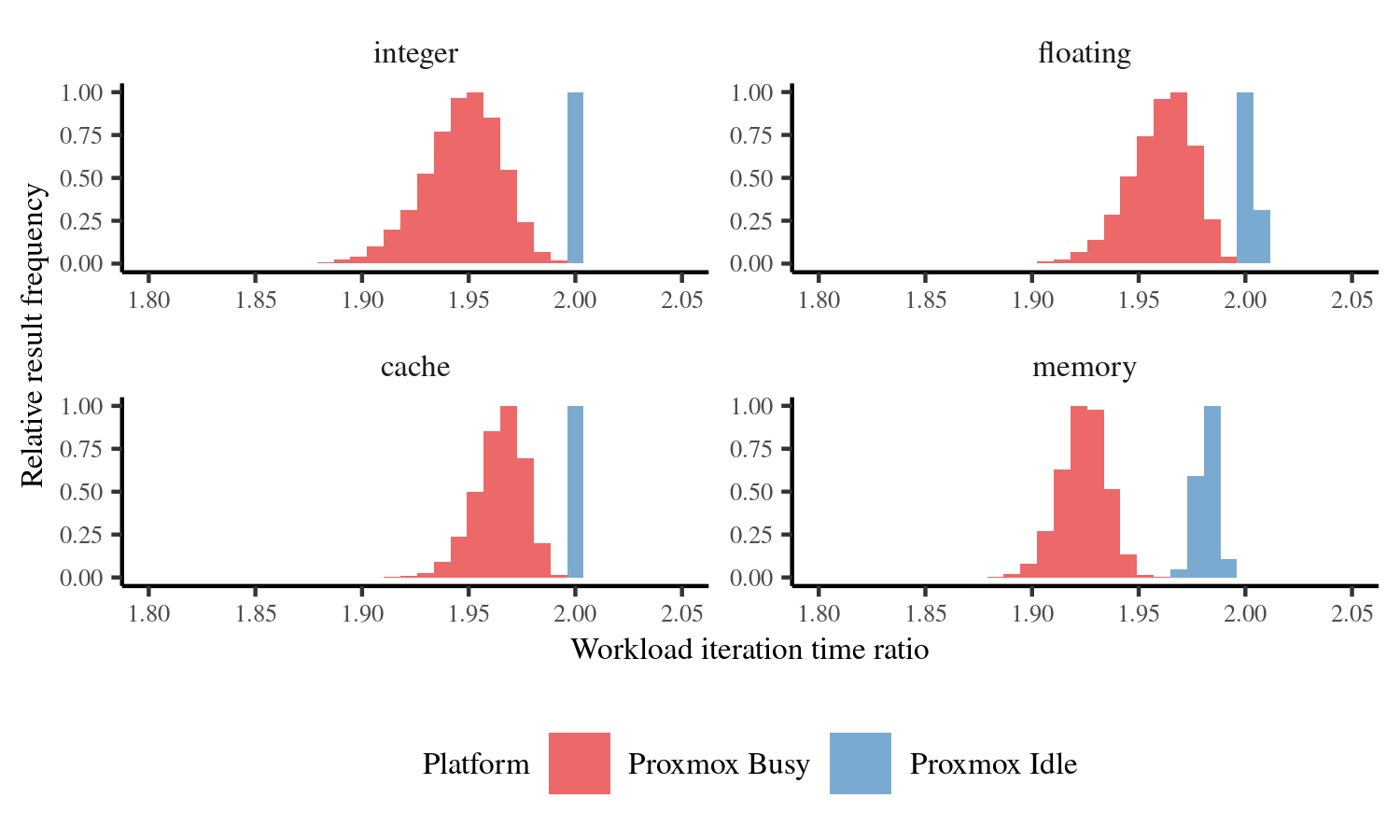}
\caption{Distribution of observed mean iteration time ratios for individual artificial workloads in private cloud, idle vs busy with competing workload, 10 runs.}
\label{fig:difference-private-cloud}
\end{figure}

As Figure~\ref{fig:difference-private-cloud} illustrates,
the observed ratio of iteration times for the two workloads is indeed very close to $2.0$.
We can observe the ratio decreasing slightly when the platform suffers from additional resource contention,
generated again using the composite configuration of the SPEC JBB 2015 benchmark running across all cores of the physical server.
This is most visible with the \emph{memory} workload, which makes practical sense because out of the four artificial workloads, \emph{memory} is most sensitive to memory bandwidth, which is shared across the entire physical server.
We can conclude that on the local cloud, the impact of uneven resource utilization is negligible.

\medskip\noindent\textbf{Public cloud.}
We can also assess the impact of uneven resource utilization using the previous A/A measurements on the public cloud.
In the private cloud, we have constructed an A/B measurement where B was twice as long as A, and examined the ratio.
Each A/B duet measurement had two phases, a concurrent phase where both A and B executed,
and an isolated phase, where A already finished and B executed in isolation.
Here, we observe that the concurrent phase of the A/B duet measurement resembles an A/A duet measurement,
and the isolated phase of the A/B duet measurement resembles a standard isolated measurement of B.
Both are measurements we have collected previously, we can therefore use the resemblance
to construct a hypothetical A/B measurement scenario.

\begin{figure*}[t!]
\includegraphics[width=\linewidth]{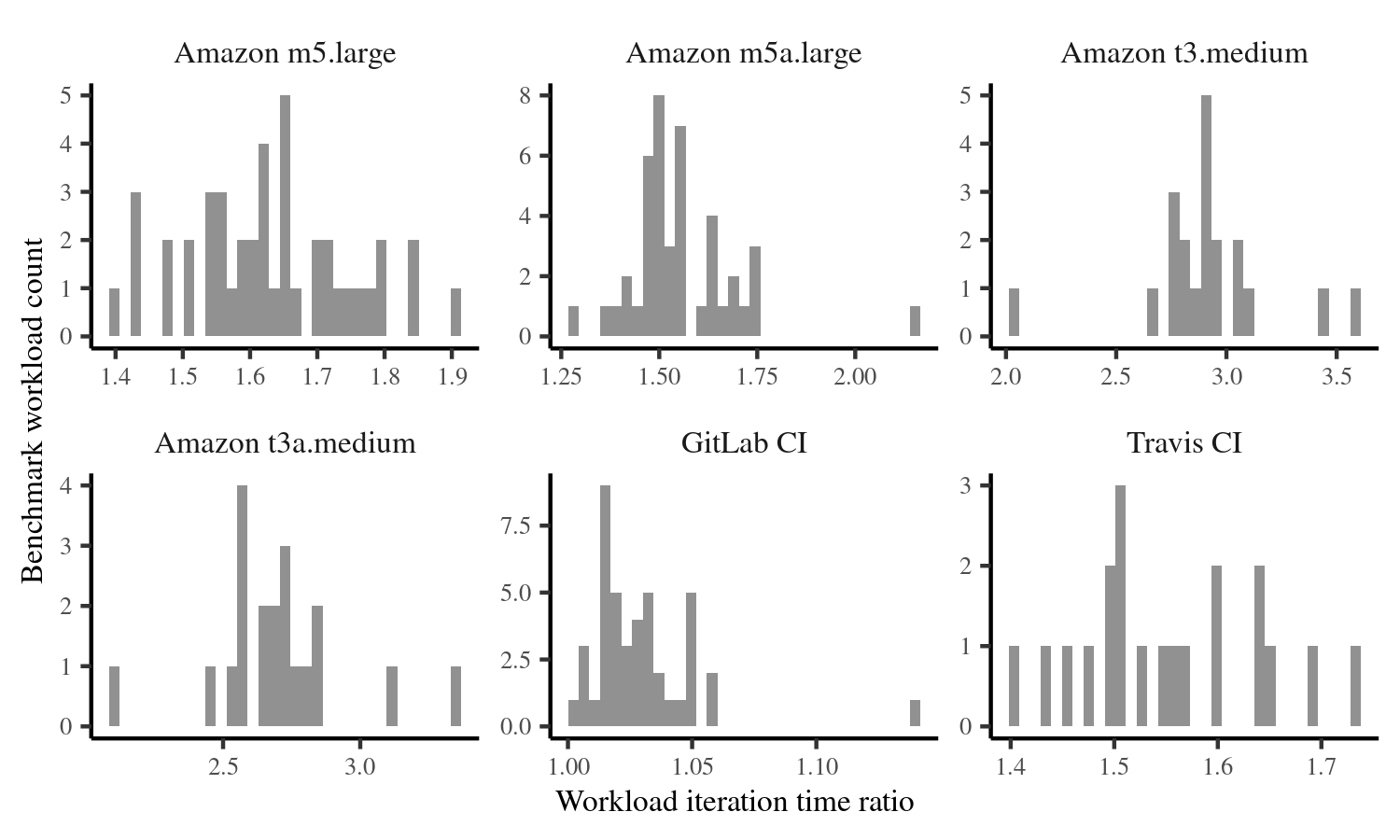}
\caption{Distribution of observed ratios of mean iteration times between A/A duet procedure measurements and standard isolated measurements.}
\label{fig:difference-public-cloud}
\end{figure*}

The ratios of mean iteration times for the public cloud platforms are in Figure~\ref{fig:difference-public-cloud}. On GitLab CI, the ratios are close to $1.0$, suggesting that the uneven resource utilization is not an issue. On the other public cloud platforms, the ratios are larger -- in other words, the same workloads take longer when executed as A/A duet measurement than when executed using a standard isolated measurement.
In the hypothetical A/B measurement scenario, this translates into an underestimated workload execution time ratio.

We attribute the difference between the platforms to two factors -- hyperthreading and token bucket processor allocation.
On the Travis CI and Amazon \instance{m} platforms, the ratios range between $1.0$ and $2.0$, which corresponds to hyperthreading splitting the computing power of a single hardware core between two virtual cores for the duet workloads.\footnote{Although the Proxmox private cloud also uses hyperthreading, it does not have the same impact. This is because the private cloud schedules virtual cores across all physical cores, unlike the Amazon public cloud, which likely binds the virtual cores to the hardware threads of one physical core.}
On the Amazon \instance{t} platforms, the ratios exceed $2.0$, likely because the token bucket processor allocation throttles the concurrent workloads executing on two virtual cores more than the isolated workloads executing on one virtual core.\footnote{Somewhat surprisingly, this would suggest that it is more cost efficient to use Amazon \instance{t} instances as single-core rather than dual-core machines.}

Returning to the research question, our results put an upper bound on how much we can underestimate the workload execution time ratio.
For example, if an A/A execution takes 3 times as much time as A executing alone, and B executing alone takes 2 times as much as A,
the desired ratio of $2.0$ would instead be measured as $(3+1)/3 \approx 1.3$. Figure~\ref{fig:difference-public-cloud} suggests
this would be an extreme case.

At the same time, our experiments provide a way to address this concern if required.
Because the underestimated workload execution time ratio is associated with uneven resource utilization,
we can simply adjust the duet procedure to continue (repeatedly) executing the shorter workload until the longer workload finishes,
rather than leaving the resources of the shorter workload idle. This measure obviously removes the uneven resource utilization.

\subsection{Discussion}

The combined answers to the four research questions prove that the duet procedure improves performance comparison accuracy on shared resource platforms by relying on the synchronized nature of resource sharing interference. Our experiments suggest the assumption of synchronized interference is safe to make on many platforms -- although it hinges on a multitude of technical details, these boil down to expecting that the platforms treat similar workloads in symmetrical situations equally.

On the flip side of the same argument, the duet procedure may not improve accuracy when comparing workloads with very different bottleneck resources, such as a CPU-bound workload and an I/O-bound workload. There is no reason to expect any resource sharing interference to impact most different resources equally. This is a threat to external validity of our results.

We can also argue that comparing workloads with different bottleneck resources is inherently fraught with issues.
The relative performance of the workloads is more likely to change between platforms with different
resource parameters, making comparison results less portable and therefore less useful.

A very general threat to both external and internal validity concerns the complex and diverse nature of public cloud platforms.
Because cloud performance characteristics may vary significantly across platforms,
our conclusions are potentially restricted to the platforms and workloads we use.
Also, some of the effects we observe may be due to internal mechanisms we do not analyze.
While characterizing every platform and workload is clearly not possible,
we do use multiple platforms and workloads to at least partially address this concern.

We have mostly limited our experiments to the application of the duet procedure for change detection in the cloud,
however, we do see more application opportunities both in the cloud and on bare metal systems.
One interesting challenge is integration into CI/CD pipelines without dedicated virtual machine instances.
Such platforms can possibly use fine-grained processor-scheduling policies in place of
binding workloads to cores, and still achieve a reasonable comparison accuracy.

\section{Related Work}
\label{sec:related}

Our related work section includes a condensed version of an earlier analysis in~\cite{BulejInitialExperiments2019}.
We start with the paper by Laaber et al.~\cite{LaaberSoftwareMicrobenchmarkingCloud2019}, which investigates the accuracy achievable in the cloud with standard measurement methods, that is, when executing the evaluated workloads one after another with randomization as recommended by~\cite{AbediConductingRepeatableExperiments2017}. Laaber et al. demonstrate that when using the standard confidence interval overlap test with 95\% confidence intervals for the mean, A/A testing needs fairly high experiment repetition counts to reduce the false alarm rate below 5\%. The authors conclude that for most of their workloads, \enquote{small slowdowns (less than 5\%) cannot reliably be detected in the cloud, at least not with the maximum number of instances (they) tested (20)}~\cite{LaaberSoftwareMicrobenchmarkingCloud2019}. Our duet procedure improves on this result.

The work of Abedi and Brecht~\cite{AbediConductingRepeatableExperiments2017} shows how the ordering of trials can impact the experiment conclusions.
Utilizing A/A testing, the authors show that possible regularity in performance interference can be incorrectly interpreted as actual difference in performance between alternatives.
Randomized ordering of trials is proposed as a remedy.
Our duet measurements similarly randomize the assignment of workloads to processors.

Existing research also often deals with the question of how many measurements to collect to achieve certain measurement accuracy, examples of recent work include He et al.~\cite{HeStatisticsBasedPerformanceTesting2019} for virtual machine instances or Maricq et al.~\cite{MaricqTamingPerformanceVariability2018} for bare metal instances.
Applying this work alongside our duet procedure is not necessarily straightforward, because the measurement accuracy metrics may not work with performance expressed as a ratio.
Other than this, the work is complementary to our duet procedure.

In a broader sense, our work is connected to research on cloud performance characteristics.
A study by Leitner and Cito~\cite{LeitnerPatternsChaosStudy2016} collects previously published observations on cloud performance and tests these observations with experiments.
Especially relevant to our work are their conclusions on the performance stability of individual instances -- this is shown to depend on the workload, with I/O-bound workload performance being sensitive to noisy neighbors, and CPU-bound workload performance depending mostly on actual allocated hardware.

Among studies that show significant performance variability in the cloud, many attribute that variability mainly to hardware heterogeneity.
Cerotti et al.~\cite{CerottiFlexibleCPUProvisioning2012} investigate the effects of hardware heterogeneity on instance performance in the Amazon public cloud, showing that instances of the same type can be backed by different CPU types and differ in performance by 20\% to 30\%.
Farley et al.~\cite{FarleyMoreYourMoney2012} also examine the effects of hardware heterogeneity in the Amazon public cloud. Different CPU types are shown to differ in performance by as much as 280\%. Differences of around 15\% are observed among different instances with the same CPU types, similar differences are observed for the same instance across time.
Ou et al.~\cite{OuSameInstanceType2013} report similar findings. For Amazon public cloud and performance differences between instances of the same type, CPU performance variability ranges between 10\% and 20\% and memory performance variability reaches as much as 270\%.
Other studies that concern various aspects of cloud performance variability include~\cite{SchadRuntimeMeasurementsCloud2010,IosupPerformanceVariabilityProduction2011,LenkWhatAreYou2011,EricsonAnalysisPerformanceVariability2017,RistovAnalysingPerformanceInstability2017,ShankarMeasuringPerformanceVariability2018}. Often, the purpose of the studies is to work towards efficient strategies of cloud resource allocation.

Although performance variability in public cloud is an accepted fact, the actual numbers observed in individual studies can rarely be compared directly due to differences in experimental settings. In our experiments, we have observed very little processor heterogeneity, and are mostly concerned with variability in time. If this were not the case, strategies to reduce processor heterogeneity in allocated instances can be utilized during testing.

Some authors propose mechanisms that help detect the presence of performance interference.
Joshi et al.~\cite{JoshiSherlockLightweightDetection2017} measure an application in controlled conditions, constructing a throughput-vs-utilization curve.
In real deployment, significant departure from that curve is interpreted as a sign of performance interference.
Similarly, Mukherjee et al.~\cite{MukherjeeSubscriberDrivenInterferenceDetection2017} measure the performance characteristics of a lightweight probe deployed together with an application, and detect interference when the workload of the application
cannot account for the changes in performance of the probe.
Both strategies require some prior knowledge of the application to be deployed, and are therefore difficult to combine with performance testing of the type we consider.

\section{Conclusion}
\label{sec:conclusion}

Our experimental evaluation on 23 SPEC CPU 2017 workloads and 20 ScalaBench and DaCapo workloads
suggests that duet measurement in the cloud is significantly more accurate than existing
methodologies based on sequential measurements.
Furthermore, our evaluation confirms that the improved accuracy
is because the paired workloads are subjected to synchronized external interference.
This external interference is an inherent property of running the workloads in the cloud,
where the underlying resources are shared with the workloads of other users --
whereas earlier techniques provide the same accuracy as duet measurement
when there is no resource sharing, their accuracy deteriorates
considerably in the presence of sharing.

The duet measurement procedure can introduce competition on the resources
between the paired workloads and uneven resource utilization patterns.
We show that these effects are either negligible or bounded and
therefore do not prevent the detection of performance regressions.

Our observations imply that duet measurement is a viable technique
for performance regression testing on both bare metal systems and
in public cloud environments that support dedicated virtual machine instances.
An interesting question is whether this technique can also improve accuracy
of CI/CD pipelines without dedicated instances---we leave the answer
to future work.

\section{Appendix}

\subsection{Experiment Configuration}

The duet measurements target shared resource environments common in clouds, most of our measurements therefore execute in clouds.
As the main cloud platform, we use the Amazon Elastic Cloud, with machine instances allocated in three different zones (us-east-1, us-east-2, us-west-2).
We use four different instance types that were the smallest general-purpose-computing instances with two virtual cores and sufficient memory -- \instance{t3.medium} (two virtual cores on Intel Xeon Platinum 8175M, reported speed $\approx$2.5 GHz 20\% burstable, 4 GB RAM), \instance{t3a.medium} (two virtual cores on AMD EPYC 7571, reported speed $\approx$2.2 GHz 20\% burstable, 4 GB RAM), \instance{m5.large} (two virtual cores on Intel Xeon Platinum 8175M, reported speed $\approx$3.1 GHz, 8 GB RAM), and \instance{m5a.large} (two virtual cores on AMD EPYC 7571, reported speed $\approx$2.8 GHz, 8 GB RAM). The instances were rented in spot mode ($\approx$0.01 USD/h for t instances, $\approx$0.04 USD/h for m instances), running Amazon Linux 2.0.20190823.

As our second cloud platform, we use the Travis CI infrastructure~\cite{softwareTravis} rented in the bootstrap plan (69 USD/month), which in turn uses otherwise unspecified Google Compute Engine platform machine instances (two virtual cores on Intel Haswell, speed 2.3 GHz burstable, 7.5 GB RAM), running Ubuntu 14.04.5 LTS.

As our third cloud platform, we use the GitLab CI infrastructure~\cite{softwareGitlabRunner} backed by Digital Ocean machine instances (two virtual cores on Intel Xeon E5-2650L V3, reported speed $\approx$1.8 GHz, 4 GB RAM), running Ubuntu 18.04.3 LTS.

In addition to the three public cloud platforms, we carry out measurements on a private cloud running the Proxmox Virtual Environment version 5.4 on multiple identical servers (Intel Xeon E3-1270 V6, 3.8 GHz, 32 GB RAM) with machine instance sizes set to roughly correspond to the public cloud instances (2 virtual cores, 8 GB RAM) and running Fedora Linux 30 with kernel 5.2.7.

Finally, we run bare metal measurements that are to represent the most stable baseline for comparison. These use multiple identical blade servers (Intel Xeon E3-1230 V6, 3.5 GHz, 32 GB RAM) with disabled hardware threads and power management features, running Fedora Linux 27 with kernel 4.15.6.

\medskip

To approximate realistic workloads, we use benchmark suites -- SPEC CPU 2017~\cite{softwareSpecCpu2017} for statically compiled and optimized workloads, and ScalaBench~\cite{SeweCapoConScala2011} (with DaCapo~\cite{BlackburnDaCapoBenchmarksJava2006}) for dynamically compiled and optimized workloads.
From SPEC CPU 2017, we execute the rate workload variants, whose system requirements and execution times match our application context better than the more demanding speed workload variants (23 workloads total).
From ScalaBench and DaCapo, we execute all workloads except actors, batik, eclipse, tomcat, tradebeans and tradesoap, which fail for various reasons ranging from known bugs to exclusive network port use that prevents concurrent execution of multiple workload instances (20 workloads total).
We use the OpenJDK 1.8.0 JVM provided by the individual Linux distributions listed above. To minimize the virtual machine startup artifacts, the JVM was run with fixed heap size (1.5 GB per JVM when the instance had 4 GB RAM, 3.5 GB per JVM when the instance had 7.5 or 8 GB RAM) and disabled garbage collector ergonomics (parallel garbage collector, young generation size 25\%, tenuring threshold 4), full garbage collection cycles were forced between individual benchmark iterations. Other virtual machine settings were left at their defaults.

\medskip

To provide information on result variance, we execute all benchmarks multiple times (with minor variations due to measurement failures and restarts, we have collected on average over 20 runs for each workload on the Amazon \instance{t} instances, over 40 runs on the Amazon m instances, and over 100 runs on the other platforms). Some virtual machine instances are used for multiple measurements. On the Amazon platform, we have an average of 20 different instances used for each workload. On GitLab CI, we have allocated 4 different instances and let the GitLab CI infrastructure choose whichever instance it decided on. With Travis CI, the allocation of instances is outside our control.
We use random samples of 10 runs for all computations.
These numbers were chosen to provide sufficient opportunity for exploring
the non deterministic execution behavior---our informal experiments suggest
that fewer runs yield too wide confidence intervals for mean iteration time.

On the relatively fast execution platforms (public cloud at full speed, private cloud, bare metal), we collect the timing of the first 100 iterations or 10 first minutes of execution within each run, whichever comes first. On the relatively slow execution platforms (public clouds with token bucket processor allocation), it is 100 iterations or 60 minutes. We do not execute the SPEC CPU 2017 workloads on the Amazon \instance{t} instances and on the Travis CI infrastructure, because both lack the computing power to execute the benchmark in reasonable time.

\medskip

For the SPEC CPU 2017 workloads, which exhibit virtually no startup artifacts, we use the timing of all iterations.
For the ScalaBench workloads, which exhibit startup artifacts related to dynamic compilation, we discard the timing of the first half of iterations as cold and only use the rest as warm. The intent is to avoid measurements taken before the compilation of the hottest methods completes, however, it is not our ambition to guarantee steady state performance measurements -- the diversity of the measurement configurations means we would have to rely on runtime steady state detection, which would introduce significant additional variability between individual executions. We have deemed it better to provide an apples-to-apples comparison on data that may include some dynamic artifacts, as opposed to an apples-to-oranges comparison on data that omitted a varying number of initial iterations.

\medskip

Finally, in all computations we employ outlier filtering with winsorization, replacing at most one observation in a run with its nearest neighbor when that observation is further than 20\% away from the min-max range of the remaining observations. Our bootstrap computations use 10000 replicates.

\subsection{Tabulated Experiment Results}

\begin{table*}[t!]
{
\centering
\begin{tabular}{lr@{$\,$:$\,$}c@{$\,$:$\,$}lr@{$\,$:$\,$}c@{$\,$:$\,$}lr@{$\,$:$\,$}c@{$\,$:$\,$}lr@{$\,$:$\,$}c@{$\,$:$\,$}l}
\toprule
Benchmark & \multicolumn{3}{c}{Amazon m5.large} & \multicolumn{3}{c}{Amazon m5a.large} & \multicolumn{3}{c}{Amazon t3.medium} & \multicolumn{3}{c}{Amazon t3a.medium} \\
\midrule
apparat & \textbf{7} & 14 & 17 & \textbf{14} & 25 & 20 & \textbf{11} & 26 & 46 & \textbf{15} & 30 & 51 \\
avrora & \textbf{1.3} & 2.5 & 2.8 & \textbf{2.8} & 12 & 12 & \textbf{4.6} & 75 & 51 & \textbf{17} & 89 & 56 \\
factorie & \textbf{7} & 11 & 12 & \textbf{9.6} & 15 & 21 & \textbf{5.7} & 12 & 41 & \textbf{17} & 22 & 55 \\
fop & \textbf{4.3} & 5.5 & 10 & \textbf{4} & 12 & 14 & \textbf{20} & 180 & 150 & \textbf{25} & 100 & 200 \\
h2 & \textbf{4.5} & 9 & 16 & \textbf{3.8} & 10 & 11 & \textbf{12} & 95 & 60 & \textbf{21} & 110 & 49 \\
jython & \textbf{1.4} & 8.5 & 9.3 & \textbf{2.3} & 12 & 12 & \textbf{3.5} & 55 & 65 & \textbf{19} & 72 & 140 \\
kiama & \textbf{3.1} & 3.8 & 3.7 & \textbf{3.9} & 11 & 13 & \textbf{10} & 97 & 61 & \textbf{48} & 72 & 110 \\
luindex & \textbf{0.96} & 5.7 & 8.7 & \textbf{1.9} & 12 & 12 & \textbf{5.6} & 110 & 54 & \textbf{16} & 30 & 31 \\
lusearch & \textbf{1.9} & 3.3 & 2.7 & \textbf{2.8} & 12 & 13 & \textbf{3.1} & 70 & 45 & \textbf{16} & 17 & 48 \\
pmd & \textbf{1.4} & 4.8 & 5.7 & \textbf{1.6} & 12 & 13 & \textbf{6.9} & 160 & 98 & \textbf{14} & 120 & 66 \\
scalac & \textbf{3.1} & 5.3 & 5.8 & \textbf{3.3} & 11 & 13 & \textbf{5.1} & 130 & 86 & \textbf{15} & 140 & 140 \\
scaladoc & \textbf{2.9} & 5.1 & 6.4 & \textbf{3.5} & 13 & 12 & \textbf{7.8} & 120 & 87 & \textbf{8.8} & 110 & 110 \\
scalap & \textbf{2.2} & 3.6 & 3.3 & \textbf{1.8} & 12 & 12 & \textbf{7.4} & 61 & 20 & \textbf{20} & 130 & 46 \\
scalariform & \textbf{3} & 4.8 & 5.2 & \textbf{3.1} & 12 & 13 & \textbf{4.3} & 150 & 100 & \textbf{12} & 130 & 100 \\
scalatest & \textbf{0.84} & 4.4 & 5.2 & \textbf{1.4} & 12 & 13 & \textbf{4} & 72 & 37 & \textbf{13} & 100 & 57 \\
scalaxb & 28 & 27 & \textbf{26} & 24 & 28 & \textbf{24} & \textbf{21} & 110 & 59 & \textbf{28} & 100 & 98 \\
specs & \textbf{4.5} & 5.1 & 6.2 & \textbf{4.9} & 12 & 11 & \textbf{6.2} & 98 & 46 & \textbf{15} & 74 & 56 \\
sunflow & \textbf{0.97} & 3.1 & 3.1 & \textbf{0.99} & 13 & 13 & \textbf{3.2} & 8.5 & 50 & \textbf{20} & 21 & 38 \\
tmt & \textbf{5.2} & 10 & 9.2 & \textbf{14} & 17 & 26 & \textbf{3} & 66 & 62 & \textbf{20} & 65 & 76 \\
xalan & \textbf{2.9} & 3.9 & 3.6 & \textbf{4} & 13 & 19 & \textbf{4.4} & 78 & 34 & \textbf{17} & 120 & 58 \\
500.perlbench\_r & \textbf{0.33} & 4.3 & 11 & \textbf{0.18} & 13 & 11 & \multicolumn{3}{c}{---} & \multicolumn{3}{c}{---} \\
502.gcc\_r & \textbf{0.096} & 5.9 & 7.4 & \textbf{0.26} & 11 & 11 & \multicolumn{3}{c}{---} & \multicolumn{3}{c}{---} \\
503.bwaves\_r & \textbf{0.023} & 2.5 & 2.1 & \textbf{0.0036} & 10 & 12 & \multicolumn{3}{c}{---} & \multicolumn{3}{c}{---} \\
505.mcf\_r & \textbf{0.11} & 4.4 & 3 & \textbf{0.21} & 9.8 & 10 & \multicolumn{3}{c}{---} & \multicolumn{3}{c}{---} \\
507.cactuBSSN\_r & \textbf{0.16} & 5.5 & 5.3 & \textbf{2} & 12 & 13 & \multicolumn{3}{c}{---} & \multicolumn{3}{c}{---} \\
508.namd\_r & \textbf{0.087} & 2.3 & 2.2 & \textbf{0.066} & 12 & 12 & \multicolumn{3}{c}{---} & \multicolumn{3}{c}{---} \\
510.parest\_r & \textbf{0.08} & 2.4 & 2.4 & \textbf{0.094} & 11 & 9.7 & \multicolumn{3}{c}{---} & \multicolumn{3}{c}{---} \\
511.povray\_r & \textbf{0.25} & 4.3 & 2 & \textbf{0.3} & 11 & 12 & \multicolumn{3}{c}{---} & \multicolumn{3}{c}{---} \\
519.lbm\_r & \textbf{0.28} & 2.5 & 3.6 & \textbf{0.28} & 8.8 & 11 & \multicolumn{3}{c}{---} & \multicolumn{3}{c}{---} \\
520.omnetpp\_r & \textbf{0.96} & 6.6 & 16 & \textbf{0.29} & 9.2 & 13 & \multicolumn{3}{c}{---} & \multicolumn{3}{c}{---} \\
521.wrf\_r & \textbf{0.017} & 2.5 & 2.4 & \textbf{0.065} & 12 & 11 & \multicolumn{3}{c}{---} & \multicolumn{3}{c}{---} \\
523.xalancbmk\_r & \textbf{0.24} & 3.8 & 4.3 & \textbf{0.22} & 10 & 11 & \multicolumn{3}{c}{---} & \multicolumn{3}{c}{---} \\
525.x264\_r & \textbf{0.045} & 2.2 & 2.3 & \textbf{0.068} & 12 & 11 & \multicolumn{3}{c}{---} & \multicolumn{3}{c}{---} \\
526.blender\_r & \textbf{0.088} & 3.8 & 3.3 & \textbf{0.13} & 12 & 13 & \multicolumn{3}{c}{---} & \multicolumn{3}{c}{---} \\
527.cam4\_r & \textbf{0.23} & 2.7 & 2.5 & \textbf{0.28} & 13 & 11 & \multicolumn{3}{c}{---} & \multicolumn{3}{c}{---} \\
531.deepsjeng\_r & \textbf{0.054} & 2.4 & 3.4 & \textbf{0.39} & 10 & 12 & \multicolumn{3}{c}{---} & \multicolumn{3}{c}{---} \\
538.imagick\_r & \textbf{0.15} & 2.7 & 2.5 & \textbf{0.42} & 11 & 11 & \multicolumn{3}{c}{---} & \multicolumn{3}{c}{---} \\
541.leela\_r & \textbf{0.19} & 4.5 & 2.2 & \textbf{0.13} & 14 & 11 & \multicolumn{3}{c}{---} & \multicolumn{3}{c}{---} \\
544.nab\_r & \textbf{0.24} & 2.9 & 2.8 & \textbf{0.089} & 12 & 11 & \multicolumn{3}{c}{---} & \multicolumn{3}{c}{---} \\
548.exchange2\_r & \textbf{0.11} & 4 & 1.9 & \textbf{0.029} & 11 & 11 & \multicolumn{3}{c}{---} & \multicolumn{3}{c}{---} \\
549.fotonik3d\_r & \textbf{0.037} & 4.7 & 2.5 & \textbf{0.04} & 8.7 & 9 & \multicolumn{3}{c}{---} & \multicolumn{3}{c}{---} \\
554.roms\_r & \textbf{0.11} & 2.8 & 2.5 & \textbf{0.36} & 10 & 13 & \multicolumn{3}{c}{---} & \multicolumn{3}{c}{---} \\
557.xz\_r & \textbf{0.6} & 4.5 & 7.2 & \textbf{0.051} & 11 & 12 & \multicolumn{3}{c}{---} & \multicolumn{3}{c}{---} \\
\bottomrule
\end{tabular}

 }
\end{table*}

\begin{table*}[t!]
{
\centering
\begin{tabular}{lr@{$\,$:$\,$}c@{$\,$:$\,$}lr@{$\,$:$\,$}c@{$\,$:$\,$}lr@{$\,$:$\,$}c@{$\,$:$\,$}lr@{$\,$:$\,$}c@{$\,$:$\,$}lr@{$\,$:$\,$}c@{$\,$:$\,$}l}
\toprule
Benchmark & \multicolumn{3}{c}{Bare Metal} & \multicolumn{3}{c}{GitLab CI} & \multicolumn{3}{c}{Proxmox Busy} & \multicolumn{3}{c}{Proxmox Idle} & \multicolumn{3}{c}{Travis CI} \\
\midrule
apparat & 18 & 16 & \textbf{16} & \textbf{22} & 75 & 79 & \textbf{17} & 33 & 29 & 16 & 18 & \textbf{15} & \textbf{12} & 22 & 18 \\
avrora & \textbf{3.9} & 4.1 & 3.9 & \textbf{2.2} & 66 & 76 & \textbf{3.3} & 13 & 11 & 4 & 4 & \textbf{3.3} & 7.3 & 11 & \textbf{7.2} \\
factorie & \textbf{16} & 18 & 19 & \textbf{14} & 49 & 52 & \textbf{13} & 29 & 23 & \textbf{14} & 14 & 17 & \textbf{8.7} & 17 & 26 \\
fop & 4.6 & \textbf{4.4} & 6.5 & \textbf{7.5} & 69 & 73 & \textbf{7.6} & 24 & 21 & 7.7 & \textbf{6.8} & 8.7 & \textbf{6.3} & 14 & 15 \\
h2 & \textbf{8.6} & 8.8 & 10 & \textbf{4.2} & 51 & 53 & \textbf{3.8} & 22 & 21 & \textbf{3.7} & 3.8 & 5.6 & \textbf{4.7} & 17 & 26 \\
jython & \textbf{6.9} & 7 & 8.6 & \textbf{5.8} & 64 & 69 & \textbf{8.6} & 29 & 27 & \textbf{7.7} & 8.2 & 12 & \textbf{2.1} & 25 & 18 \\
kiama & 3.3 & \textbf{3.1} & 3.3 & \textbf{4.1} & 63 & 69 & \textbf{2.6} & 24 & 21 & 3.1 & 2.9 & \textbf{2.8} & \textbf{2.3} & 13 & 20 \\
luindex & \textbf{0.79} & 8.1 & 5.6 & \textbf{3.5} & 60 & 63 & \textbf{2} & 16 & 15 & \textbf{2.8} & 7.7 & 5.3 & \textbf{1.7} & 11 & 10 \\
lusearch & \textbf{2.3} & 2.5 & 2.6 & \textbf{2.5} & 61 & 63 & \textbf{2.1} & 19 & 15 & \textbf{2.1} & 2.4 & 2.5 & \textbf{1.9} & 20 & 25 \\
pmd & 2.1 & 1.8 & \textbf{1.7} & \textbf{3.7} & 69 & 70 & \textbf{1.4} & 23 & 22 & 1.6 & \textbf{1.5} & 1.6 & \textbf{1.1} & 16 & 11 \\
scalac & 5.6 & \textbf{5.3} & 5.6 & \textbf{5} & 66 & 66 & \textbf{4.6} & 21 & 19 & 4.8 & \textbf{4.5} & 4.6 & \textbf{6} & 13 & 15 \\
scaladoc & \textbf{4.3} & 4.4 & 5 & \textbf{4.4} & 64 & 69 & \textbf{3.8} & 23 & 20 & 4.2 & 4.3 & \textbf{4.1} & \textbf{2.7} & 10 & 21 \\
scalap & \textbf{2.1} & 3.1 & 2.4 & \textbf{5} & 56 & 58 & \textbf{2.2} & 17 & 14 & \textbf{2.6} & 2.8 & 2.9 & \textbf{2.6} & 9.2 & 11 \\
scalariform & \textbf{4.2} & 4.3 & 5.2 & \textbf{5.3} & 59 & 67 & \textbf{2.8} & 19 & 16 & \textbf{3.1} & 3.1 & 4 & \textbf{4.1} & 10 & 15 \\
scalatest & \textbf{1.5} & 1.5 & 2.3 & \textbf{3.1} & 69 & 67 & \textbf{1.3} & 21 & 16 & 1.5 & \textbf{1.5} & 1.7 & \textbf{1.4} & 10 & 12 \\
scalaxb & 30 & 27 & \textbf{20} & \textbf{23} & 70 & 74 & \textbf{28} & 37 & 32 & 29 & 26 & \textbf{19} & \textbf{23} & 27 & 27 \\
specs & \textbf{2.9} & 2.9 & 4.2 & \textbf{3.8} & 60 & 62 & \textbf{4.6} & 18 & 15 & 4.6 & \textbf{4.6} & 5.5 & \textbf{6.7} & 10 & 19 \\
sunflow & \textbf{1.9} & 2.3 & 2.2 & \textbf{2.8} & 55 & 60 & \textbf{2.3} & 25 & 20 & \textbf{2.7} & 2.8 & 3.3 & \textbf{1.1} & 11 & 7.7 \\
tmt & 5.1 & \textbf{4.8} & 18 & \textbf{12} & 61 & 71 & \textbf{7.7} & 24 & 22 & \textbf{7.1} & 7.3 & 16 & \textbf{13} & 21 & 18 \\
xalan & \textbf{2.3} & 2.4 & 2.3 & \textbf{3.5} & 73 & 74 & \textbf{2.6} & 18 & 15 & 2.9 & 2.7 & \textbf{2.5} & \textbf{3.3} & 12 & 11 \\
500.perlbench\_r & 0.71 & 0.65 & \textbf{0.57} & \textbf{2.4} & 43 & 46 & \textbf{0.42} & 28 & 30 & 0.62 & 0.55 & \textbf{0.48} & \multicolumn{3}{c}{---} \\
502.gcc\_r & \textbf{0.59} & 0.61 & 0.77 & \textbf{0.81} & 42 & 41 & \textbf{0.24} & 23 & 23 & \textbf{0.45} & 0.57 & 0.75 & \multicolumn{3}{c}{---} \\
503.bwaves\_r & \textbf{0.19} & 0.35 & 0.35 & \textbf{1.9} & 61 & 74 & \textbf{0.28} & 16 & 15 & \textbf{0.42} & 0.54 & 0.52 & \multicolumn{3}{c}{---} \\
505.mcf\_r & 1.3 & \textbf{1.1} & 1.3 & \textbf{1.9} & 41 & 42 & \textbf{0.64} & 19 & 16 & 1.3 & 1.2 & \textbf{1} & \multicolumn{3}{c}{---} \\
507.cactuBSSN\_r & 5 & 3.8 & \textbf{0.53} & \textbf{0.96} & 33 & 31 & \textbf{1.1} & 21 & 21 & 0.67 & 0.65 & \textbf{0.57} & \multicolumn{3}{c}{---} \\
508.namd\_r & \textbf{0.69} & 0.74 & 0.91 & \textbf{1.3} & 41 & 49 & \textbf{0.45} & 27 & 25 & \textbf{0.55} & 0.61 & 0.68 & \multicolumn{3}{c}{---} \\
510.parest\_r & \textbf{0.21} & 0.29 & 0.31 & \textbf{1.8} & 49 & 60 & \textbf{0.2} & 17 & 16 & \textbf{0.22} & 0.29 & 0.34 & \multicolumn{3}{c}{---} \\
511.povray\_r & 0.91 & 0.93 & \textbf{0.84} & \textbf{3.6} & 51 & 49 & \textbf{0.91} & 27 & 25 & 1 & 0.94 & \textbf{0.79} & \multicolumn{3}{c}{---} \\
519.lbm\_r & 0.36 & 0.91 & \textbf{0.24} & \textbf{1.5} & 41 & 39 & \textbf{1.2} & 18 & 17 & 1.9 & 2.1 & \textbf{1} & \multicolumn{3}{c}{---} \\
520.omnetpp\_r & 1.1 & \textbf{1.1} & 1.1 & \textbf{2.9} & 39 & 36 & \textbf{0.61} & 19 & 19 & 0.92 & \textbf{0.91} & 1.1 & \multicolumn{3}{c}{---} \\
521.wrf\_r & 0.38 & 0.68 & \textbf{0.26} & \textbf{1.2} & 48 & 52 & \textbf{0.43} & 20 & 19 & \textbf{0.37} & 0.48 & 0.65 & \multicolumn{3}{c}{---} \\
523.xalancbmk\_r & 1.9 & \textbf{1.7} & 2 & \textbf{2.4} & 86 & 97 & \textbf{0.74} & 17 & 16 & 1.5 & 1.4 & \textbf{1.3} & \multicolumn{3}{c}{---} \\
525.x264\_r & \textbf{0.15} & 0.27 & 0.24 & \textbf{3} & 46 & 50 & \textbf{1.7} & 20 & 18 & 0.95 & 2.7 & \textbf{0.17} & \multicolumn{3}{c}{---} \\
526.blender\_r & 0.33 & \textbf{0.31} & 0.33 & \textbf{2.3} & 41 & 42 & \textbf{0.29} & 24 & 20 & 0.36 & 0.39 & \textbf{0.35} & \multicolumn{3}{c}{---} \\
527.cam4\_r & \textbf{0.28} & 0.32 & 0.38 & \textbf{1.5} & 39 & 53 & \textbf{0.35} & 21 & 17 & \textbf{0.5} & 0.63 & 0.61 & \multicolumn{3}{c}{---} \\
531.deepsjeng\_r & \textbf{0.17} & 0.17 & 0.18 & \textbf{2.3} & 44 & 45 & \textbf{0.19} & 22 & 22 & \textbf{0.25} & 0.26 & 0.29 & \multicolumn{3}{c}{---} \\
538.imagick\_r & 1.6 & 0.87 & \textbf{0.12} & \textbf{2.2} & 54 & 56 & \textbf{0.56} & 26 & 27 & 0.86 & \textbf{0.8} & 0.92 & \multicolumn{3}{c}{---} \\
541.leela\_r & 0.13 & 0.14 & \textbf{0.12} & \textbf{2.6} & 37 & 40 & \textbf{0.16} & 18 & 17 & \textbf{0.14} & 0.2 & 0.18 & \multicolumn{3}{c}{---} \\
544.nab\_r & 0.99 & 1 & \textbf{0.89} & \textbf{2.2} & 54 & 60 & \textbf{0.16} & 15 & 12 & \textbf{0.36} & 0.36 & 0.71 & \multicolumn{3}{c}{---} \\
548.exchange2\_r & 0.79 & \textbf{0.72} & 0.72 & \textbf{3.8} & 49 & 58 & \textbf{0.35} & 27 & 24 & 0.46 & 0.44 & \textbf{0.37} & \multicolumn{3}{c}{---} \\
549.fotonik3d\_r & \textbf{0.026} & 1.6 & 0.65 & \textbf{3.2} & 29 & 32 & \textbf{0.099} & 18 & 23 & \textbf{0.068} & 1.3 & 0.92 & \multicolumn{3}{c}{---} \\
554.roms\_r & \textbf{0.22} & 1.1 & 0.45 & \textbf{1.9} & 37 & 41 & \textbf{0.28} & 20 & 20 & \textbf{0.88} & 1.6 & 1.1 & \multicolumn{3}{c}{---} \\
557.xz\_r & 0.54 & 0.47 & \textbf{0.47} & \textbf{1.4} & 36 & 37 & \textbf{0.31} & 19 & 17 & 0.58 & \textbf{0.56} & 0.61 & \multicolumn{3}{c}{---} \\
\bottomrule
\end{tabular}

 }
\end{table*}

For insight beyond the aggregate figures, we include complete relative confidence interval widths for all workload and platform combinations presented.
The columns list triplets separated by colons:

\begin{itemize}
    \item The interval width computed with the duet procedure.
    \item The interval width computed with the duet procedure after shuffling.
    \item The interval width computed with the standard method.
\end{itemize}

All widths are relative and expressed in percents.
The best (smallest) interval width is shown in boldface.
The listing also includes bare metal measurements, which
show the accuracy achievable under ideally controlled conditions.

\bibliographystyle{plain}
\bibliography{main}

\end{document}